\documentclass[twocolumn,twoside]{revtex4-1}
\pdfoutput=1
\usepackage{graphicx}
\usepackage{fancyhdr}
\usepackage{ifthen} 
\newboolean{uprightparticles}
\setboolean{uprightparticles}{false} 
\usepackage{xspace} 
\usepackage{upgreek}

\newcommand{\offsetoverline}[2][0.1em]{\kern #1\overline{\kern -#1 #2}}%

\def\lhcb   {\mbox{LHCb}\xspace}

\def\lhc    {\mbox{LHC}\xspace}

\def\MagUp {\mbox{\em Mag\kern -0.05em Up}\xspace}

\ifthenelse{\boolean{uprightparticles}}%
{

 \def\Pmu         {\ensuremath{\upmu}\xspace}

 \def\Ppi         {\ensuremath{\uppi}\xspace}

 \def\PDelta      {\ensuremath{\Delta}\xspace}                 
 \def\PXi         {\ensuremath{\Xi}\xspace}                 
 \def\PLambda     {\ensuremath{\Lambda}\xspace}                 
 \def\PSigma      {\ensuremath{\Sigma}\xspace}                 
 \def\POmega      {\ensuremath{\Omega}\xspace}                 
 \def\PUpsilon    {\ensuremath{\Upsilon}\xspace}

 \def\PB      {\ensuremath{\mathrm{B}}\xspace}                 
                  
 \def\PD      {\ensuremath{\mathrm{D}}\xspace}

 \def\PK      {\ensuremath{\mathrm{K}}\xspace}

 \def\Pb      {\ensuremath{\mathrm{b}}\xspace}                 
 \def\Pc      {\ensuremath{\mathrm{c}}\xspace}

 \def\Pi      {\ensuremath{\mathrm{i}}\xspace}

 \def\Pp      {\ensuremath{\mathrm{p}}\xspace}

 \def\Ps      {\ensuremath{\mathrm{s}}\xspace}                 
                  
 \def\Pu      {\ensuremath{\mathrm{u}}\xspace}

}
{

 \def\Pmu         {\ensuremath{\mu}\xspace}

 \def\Ppi         {\ensuremath{\pi}\xspace}

 \mathchardef\PDelta="7101
 \mathchardef\PXi="7104
 \mathchardef\PLambda="7103
 \mathchardef\PSigma="7106
 \mathchardef\POmega="710A
 \mathchardef\PUpsilon="7107
                  
 \def\PB      {\ensuremath{B}\xspace}                 
                  
 \def\PD      {\ensuremath{D}\xspace}

 \def\PK      {\ensuremath{K}\xspace}

 \def\Pb      {\ensuremath{b}\xspace}                 
 \def\Pc      {\ensuremath{c}\xspace}

 \def\Pi      {\ensuremath{i}\xspace}

 \def\Pp      {\ensuremath{p}\xspace}

 \def\Ps      {\ensuremath{s}\xspace}                 
                  
 \def\Pu      {\ensuremath{u}\xspace}

}

\makeatletter
\newcommand{\miniscule}{\@setfontsize\miniscule{4}{5}}
\makeatother

\DeclareRobustCommand{\optbar}[1]{\shortstack{{\miniscule (\rule[.5ex]{1.25em}{.18mm})}
  \\ [-.7ex] $#1$}}

\def\mun        {{\ensuremath{\Pmu^-}}\xspace} 

\def\uquark    {{\ensuremath{\Pu}}\xspace}

\def\squark    {{\ensuremath{\Ps}}\xspace}

\def\cquark    {{\ensuremath{\Pc}}\xspace}
\def\cquarkbar {{\ensuremath{\overline \cquark}}\xspace}

\def\bquark    {{\ensuremath{\Pb}}\xspace}

\def\pion   {{\ensuremath{\Ppi}}\xspace}

\def\pip    {{\ensuremath{\pion^+}}\xspace}
\def\pim    {{\ensuremath{\pion^-}}\xspace}
\def\pipm   {{\ensuremath{\pion^\pm}}\xspace}
\def\pimp   {{\ensuremath{\pion^\mp}}\xspace}

\def\kaon    {{\ensuremath{\PK}}\xspace}
  \def\Kbar    {{\kern 0.2em\overline{\kern -0.2em \PK}{}}\xspace}

\def\KorKbar {\kern 0.18em\optbar{\kern -0.18em K}{}\xspace}

\def\Kp      {{\ensuremath{\kaon^+}}\xspace}
\def\Km      {{\ensuremath{\kaon^-}}\xspace}
\def\Kpm     {{\ensuremath{\kaon^\pm}}\xspace}

\def\KS      {{\ensuremath{\kaon^0_{\mathrm{S}}}}\xspace}

  \def\Dbar    {{\kern 0.2em\overline{\kern -0.2em \PD}{}}\xspace}
\def\D       {{\ensuremath{\PD}}\xspace}

\def\DorDbar {\kern 0.18em\optbar{\kern -0.18em D}{}\xspace}
\def\Dz      {{\ensuremath{\D^0}}\xspace}
\def\Dzb     {{\ensuremath{\Dbar{}^0}}\xspace}

\def\Dstarp  {{\ensuremath{\D^{*+}}}\xspace}

\def\theDstarp{{\ensuremath{\D^{*}(2010)^{+}}}\xspace}

\def\B       {{\ensuremath{\PB}}\xspace}
\def\Bbar    {{\ensuremath{\kern 0.18em\overline{\kern -0.18em \PB}{}}}\xspace}

\def\BorBbar    {\kern 0.18em\optbar{\kern -0.18em B}{}\xspace}

\def\Y#1S{\ensuremath{\PUpsilon{(#1S)}}\xspace}

\def\proton      {{\ensuremath{\Pp}}\xspace}

\def\LorLbar     {\kern 0.18em\optbar{\kern -0.18em \PLambda}{}\xspace}

\newcommand{\decay}[2]{\mbox{\ensuremath{#1\!\to #2}}\xspace}         

\def\to                 {\ensuremath{\rightarrow}\xspace}

\def\CP                {{\ensuremath{C\!P}}\xspace}
\def\CPV               {{\ensuremath{C\!PV}}\xspace}

\def\Vus  {{\ensuremath{V_{\uquark\squark}}}\xspace}
\def\Vcs  {{\ensuremath{V_{\cquark\squark}}}\xspace}

\def\Vub  {{\ensuremath{V_{\uquark\bquark}}}\xspace}
\def\Vcb  {{\ensuremath{V_{\cquark\bquark}}}\xspace}

\def\AT#1     {\ensuremath{A_{\mathrm{T}}^{#1}}\xspace}           

\def\C#1      {\ensuremath{\mathcal{C}_{#1}}\xspace}                       
\def\Cp#1     {\ensuremath{\mathcal{C}_{#1}^{'}}\xspace}                    
\def\Ceff#1   {\ensuremath{\mathcal{C}_{#1}^{\mathrm{(eff)}}}\xspace}        
\def\Cpeff#1  {\ensuremath{\mathcal{C}_{#1}^{'\mathrm{(eff)}}}\xspace}       
\def\Ope#1    {\ensuremath{\mathcal{O}_{#1}}\xspace}                       
\def\Opep#1   {\ensuremath{\mathcal{O}_{#1}^{'}}\xspace}                    

\def\agamma     {\ensuremath{A_{\Gamma}}\xspace}

\newcommand{\ket}[1]{\ensuremath{|#1\rangle}}              

\newcommand{\aunit}[1]{\ensuremath{\text{\,#1}}}       

\newcommand{\tev}{\aunit{Te\kern -0.1em V}\xspace}
\newcommand{\gev}{\aunit{Ge\kern -0.1em V}\xspace}
\newcommand{\mev}{\aunit{Me\kern -0.1em V}\xspace}
\newcommand{\kev}{\aunit{ke\kern -0.1em V}\xspace}
\newcommand{\ev}{\aunit{e\kern -0.1em V}\xspace}
\newcommand{\mevc}{\ensuremath{\aunit{Me\kern -0.1em V\!/}c}\xspace}
\newcommand{\gevc}{\ensuremath{\aunit{Ge\kern -0.1em V\!/}c}\xspace}
\newcommand{\mevcc}{\ensuremath{\aunit{Me\kern -0.1em V\!/}c^2}\xspace}
\newcommand{\gevcc}{\ensuremath{\aunit{Ge\kern -0.1em V\!/}c^2}\xspace}

\def\fb   {\ensuremath{\aunit{fb}}\xspace}
\def\invfb   {\ensuremath{\fb^{-1}}\xspace}

\def\ps   {\ensuremath{\aunit{ps}}\xspace}

\def\gsim{{~\raise.15em\hbox{$>$}\kern-.85em
          \lower.35em\hbox{$\sim$}~}\xspace}
\def\lsim{{~\raise.15em\hbox{$<$}\kern-.85em
          \lower.35em\hbox{$\sim$}~}\xspace}

\def\sqs   {\ensuremath{\protect\sqrt{s}}\xspace}

\xspace

\def\tell1  {TELL1\xspace}
\def\ukl1   {UKL1\xspace}

\usepackage{amsmath}
\usepackage{physics}
\usepackage[pdftex,
            pdfauthor={Tommaso Pajero},
            pdftitle={Measurements of time-dependent CP violation and mixing in charm at LHCb},
            pdfkeywords={flavour physics, charm physics, CP violation, oscillation}
            ]{hyperref}

\begin{document}

\title{Measurements of time-dependent $\vb*{CP}$ violation and mixing in charm at \lhcb}

%

\author{T. Pajero                           \\
        on behalf of the \lhcb~collaboration}
\affiliation{Scuola Normale Superiore and INFN Sezione di Pisa, Pisa, Italy}

\begin{abstract}
The \lhcb experiment has opened the possibility to test mixing and \CP violation in the charm sector with unprecedented precision thanks to the huge number of charm hadron decays collected, $\mathcal{O}(10^8)$.
The first observation of \CP violation in the decay of charm quarks in March 2019 has been a fundamental achievement.
The latest \lhcb measurements in the complementary sectors of mixing and time-dependent \CP violation are illustrated in these proceedings.
In particular, a new measurement of the \CP violation parameter \agamma with 2015--2016 data that was presented for the first time at this conference is described.
In the last section, prospects are given for the improvements in precision expected in the next few years.
\end{abstract}

\maketitle

\thispagestyle{fancy}

\section{Introduction}
The charm-quark sector offers a unique opportunity to test the Cabibbo--Kobayashi--Maskawa (CKM) paradigm of \CP violation (\CPV)~\cite{CKM}, since it provides access to operators that affect only up-type quarks, while leaving the \kaon and \B mesons possible unaffected.
However, the smallness of the elements of the CKM matrix
involved
and the Glashow--Iliopoulos--Maiani mechanism suppress the expectations for \CPV in charm at a level typically below $10^{-3}$~\cite{Grossman}.
Therefore, testing the Standard Model (SM) expectations for \CPV in charm requires huge data samples, $\mathcal{O}(10^{7})$ decays, that have become available only recently thanks to the large \cquark\cquarkbar production cross-section at the \lhc~\cite{charm-cross-section} and to the dedicated detector and trigger of the \lhcb experiment~\cite{LHCb-detector}.
This has set the \lhcb experiment as the main player in this quest.

The \lhcb~collaboration announced the first observation of \CPV in the decay of charm quarks in March 2019~\cite{LHCb-PAPER-2019-006}.
However, the interpretation of this observation within the paradigm of the SM is controversial, since precise predictions are made difficult by low-energy quantum-chromodynamics effects~\cite{deltaACP1,deltaACP2,deltaACP3,deltaACP4}.
For this reason, further studies of charm decays are needed to clarify the picture.
Measurements of mixing and time-dependent \CPV in neutral charm mesons provide a test of the SM complementary to the measurements of \CPV in the decay and might help in this regard.
The most recent results from the \lhcb~collaboration in these two fields are presented in Sects.~\ref{sect:D02KsPP} and \ref{sect:agamma}, respectively.
Section~\ref{sect:conclusions} concludes summarising the prospects for the improvements in precision which are expected in the next few years.

\section{Measurement of the mass difference between neutral charm-meson eigenstates with $\boldmath{\decay{\Dz}{\KS\pip\pim}}$ decays \label{sect:D02KsPP}}
The split of the masses ($m_{1,2}$) and of the decay widths ($\Gamma_{1,2}$) of the neutral-charm-meson eigenstates $\ket{\D_{1,2}} \equiv p \ket{\Dz} \pm q \ket{\Dzb}$, with $\lvert p \rvert^2 + \lvert q \rvert^2 = 1$, governs the oscillations of charm neutral mesons and can be conveniently parametrised through the mixing parameters $x \equiv 2(m_2 - m_1) / (\Gamma_1 + \Gamma_2)$ and $y \equiv (\Gamma_2 - \Gamma_1) / (\Gamma_1 + \Gamma_2)$.
While the measurement of mixing with the two-body decays \decay{\Dz}{\Kpm\pimp} led to the discovery of mixing in charm and provides the most precise measurement of the parameter $y$~\cite{LHCb-PAPER-2017-046}, it supplies only limited information on the mixing parameter $x$, owing to the smallness of the difference between the strong phases of \decay{\Dz}{\Kp\pim} and \decay{\Dzb}{\Kp\pim} decays and to the large uncertainty with which this difference is measured~\cite{strong-phase}.
On the contrary, the rich resonance spectrum of \decay{\Dz}{\KS\pip\pim} decays implies the presence of large strong phases that vary across the Dalitz plane and, consequently, provides good sensitivity to all mixing and time-dependent-\CPV parameters.
However, the decay dynamics of this three-body decay and the variations of the detector efficiency across the Dalitz plane as a function of decay time need to be modelled carefully in order to take advantage of this feature.
This is expected to become more and more difficult as the size of the data samples increases, especially at a hadron collider like the \lhc where the impact of the trigger on the acceptance is nontrivial.

\begin{figure}[bt]
    \includegraphics[width=70mm]{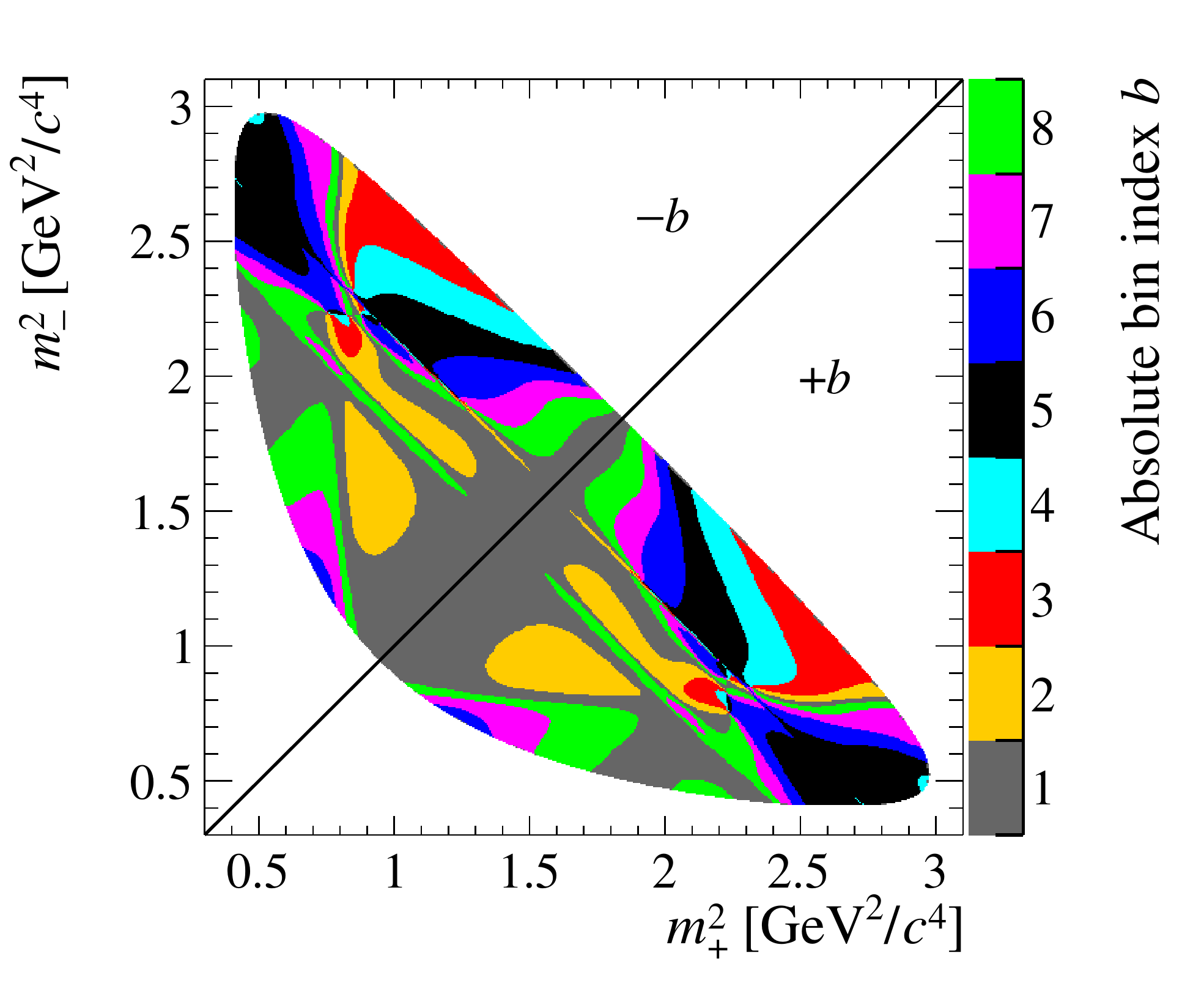} \\
    \vspace*{-0.2cm}
    \includegraphics[width=70mm]{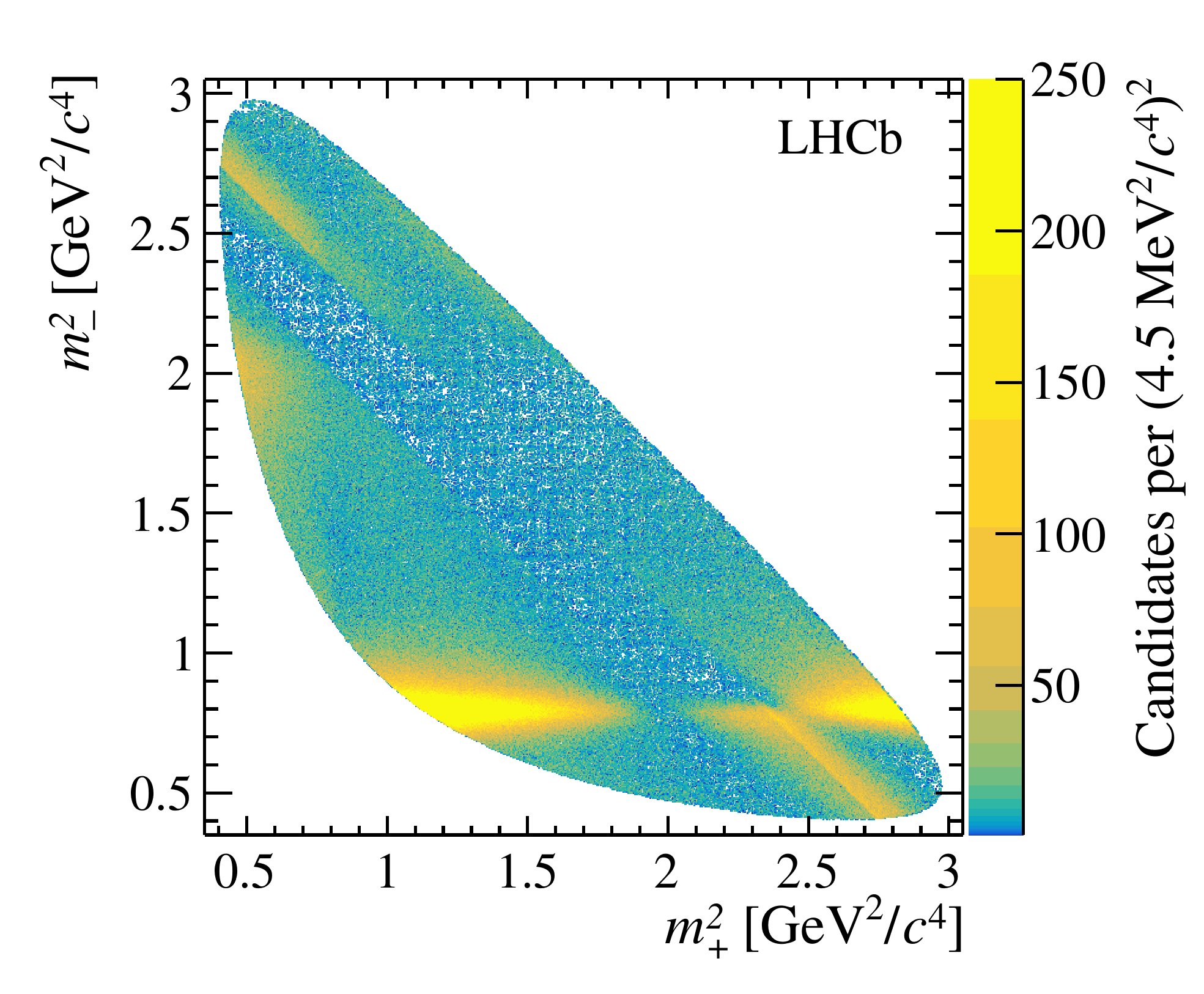}
    \vspace*{-0.5cm}
    \caption{
            (Top) iso-$\Delta\delta$ binning of the \decay{\Dz}{\KS\pip\pim} Dalitz plot, reproduced from Ref.~\cite{CLEO}.
            (Bottom) Dalitz-plot distribution of background-subtracted \decay{\Dz}{\KS\pip\pim} candidates in Ref.~\cite{LHCb-PAPER-2019-001}.}
    \label{fig:D02KsPP_binning}
\end{figure}

Both these challenges are mitigated by the ``bin-flip method'' proposed in Ref.~\cite{binflip}, a model-independent analysis procedure optimised for the measurement of the parameter $x$.
This consists in dividing the Dalitz plane into two sets of regions, symmetrically distributed with respect to its bisector $m^2_+ = m^2_-$, where $m^2_\pm$ is equal to $m^2(\KS\pipm)$ for \Dz decays and to $m^2(\KS\pimp)$ for \Dzb decays,
chosen so as to keep the strong phase difference ($\Delta\delta$) between \Dz and \Dzb decays approximately constant within each region.
Eight regions are currently used for each half of the plane, as displayed in Fig.~\ref{fig:D02KsPP_binning} (top).
For each Dalitz bin, the data sample is then further divided into bins of decay time.
Finally, the ratio of the decay yields in the pairs of bins symmetric with respect to the bisector of the Dalitz plane is considered as a function of decay time, separately for \Dz and \Dzb candidates.
Most acceptance and efficiency effects cancel in the ratio, greatly reducing the systematic uncertainties due to the limited precision with which they are known.
However, the time-dependent ratios are still sensitive to mixing.
In fact, the lower part of the Dalitz plane ($m^2_+ > m^2_-$) is dominated by unmixed, Cabibbo-favoured \Dz decays, see Fig.~\ref{fig:D02KsPP_binning} (bottom), while in the upper part of the Dalitz plane ($m^2_+ < m^2_-$) the contribution of Cabibbo-favoured decays following mixing becomes more and more important with respect to the unmixed doubly-Cabibbo-suppressed decays as decay time increases.
Therefore, these ratios are nontrivial functions of all the mixing and \CPV parameters, of the average strong phase difference between the two bins, of the average decay time and of the average squared decay time in the bins.

This method is employed in the recent \lhcb measurement of mixing and \CPV with \decay{\Dz}{\KS\pip\pim} decays using the 2011--2012 data sample, corresponding to an integrated luminosity of $1(2)\invfb$ of \proton\proton collisions at $\sqs = 7(8)\tev$~\cite{LHCb-PAPER-2019-001}.
The flavour at production of the \Dz meson is inferred either from the charge of the accompanying pion in \decay{\theDstarp}{\Dz\pip} decays, where the \theDstarp (hereafter referred to as \Dstarp) is produced in the \proton\proton collision vertex, or from the charge of the muon in \decay{\Bbar}{\Dz\mun X} decays, where $X$ indicates an arbitrary number of unreconstructed particles.
The two data samples correspond to 1.3 and 1.0 million decays, respectively.
Detector-induced variations of the efficiency in the plane of decay time \textit{vs.}~$m^2(\pip\pim)$, which might bias the results, are removed by assigning per-candidate weights proportional to the inverse of the relative efficiency, as determined through a data-driven method.
Then, the decay yield is measured, separately in each bin of the Dalitz plane and of decay time and for \Dz and \Dzb candidates, through a fit to the $m(\KS\pip\pim\pip) - m(\KS\pip\pim)$ distribution for the \pip-tagged sample or to the $m(\KS\pip\pim)$ distribution for the \mun-tagged sample.
Finally, the mixing and \CPV parameters are measured through a least-squares fit to the time-dependent ratios of the yields in the bins symmetric with respect to the bisector of the Dalitz plane, simultaneously for all Dalitz bins, for the \Dz and \Dzb candidates and for the \pip- and \mun-tagged samples.
In the fit, the strong phase differences are constrained to the values measured by CLEO~\cite{CLEO} through a Gaussian penalty term and the mixing and \CPV parameters are parametrised through the two complex parameters defined as $z_\CP \pm \Delta z \equiv - (q/p)^{\pm 1} (y + ix)$~\cite{binflip}.
The results are
\begin{alignat*}{2}
x_\CP    &\equiv -\text{Im}(z_\CP)    &= (\phantom{-}2.7\phantom{3} \pm 1.6\phantom{0} \pm 0.4\phantom{2}) \times 10^{-3}, \\
\Delta x &\equiv -\text{Im}(\Delta z) &= (          -0.53           \pm 0.70           \pm 0.22          ) \times 10^{-3}, \\  
y_\CP    &\equiv -\text{Re}(z_\CP)    &= (\phantom{-}7.4\phantom{3} \pm 3.6\phantom{0} \pm 1.1\phantom{2}) \times 10^{-3}, \\  
\Delta y &\equiv -\text{Re}(\Delta z) &= (\phantom{-}0.6\phantom{3} \pm 1.6\phantom{0} \pm 0.3\phantom{2}) \times 10^{-3},
\end{alignat*}
where the first uncertainties are statistical and the second systematic (see Ref.~\cite{binflip} for the explicit expressions of $x_\CP$, $\Delta x$, $y_\CP$ and $\Delta y = A_\Gamma$ as a function of $x$, $y$, $\lvert q/p \rvert$ and $\phi \equiv \arg(-q/p)$).
The systematic uncertainty on $x_\CP$ is dominated by a 3\% contamination of secondary \Dstarp decays, where the \Dstarp comes from the decay of a \bquark hadron instead of from the \proton\proton vertex, for the \pip-tagged sample, and by a 1\% contamination of random associations of \Dz mesons with unrelated muons for the \mun-tagged sample.
For $y_\CP$, the dominant systematic uncertainties come from neglecting the decay-time and $m^2_\pm$ resolutions and from neglecting the efficiency variations across decay time and Dalitz plot.
Finally, asymmetric nonuniformities of the reconstruction efficiency with respect to the bisector of the Dalitz plot are responsible for most of the systematic uncertainty on $\Delta x$ and $\Delta y$.

\begin{figure}[bt]
    \includegraphics[width=70mm]{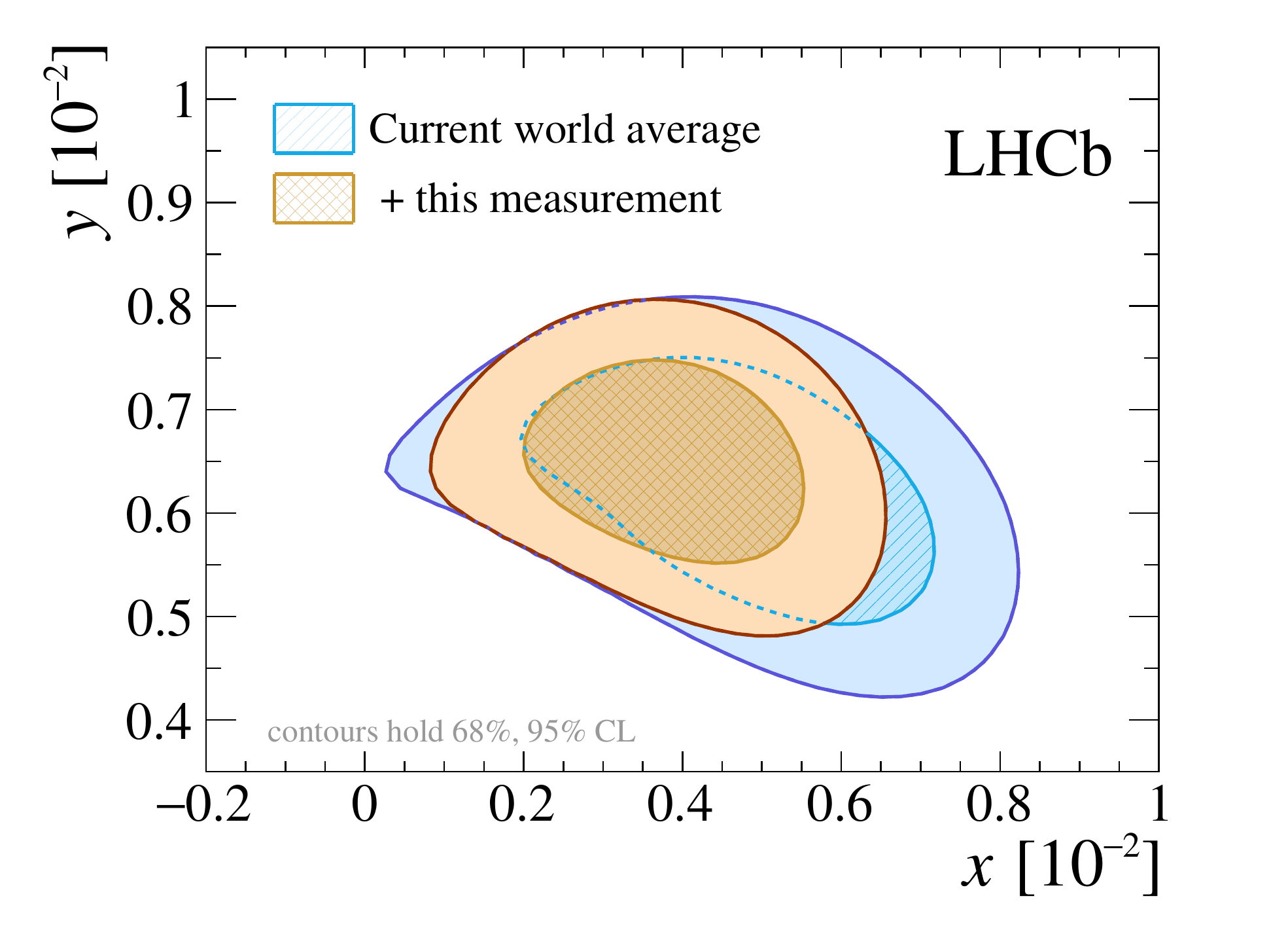} \\
    \vspace*{-0.2cm}
    \includegraphics[width=70mm]{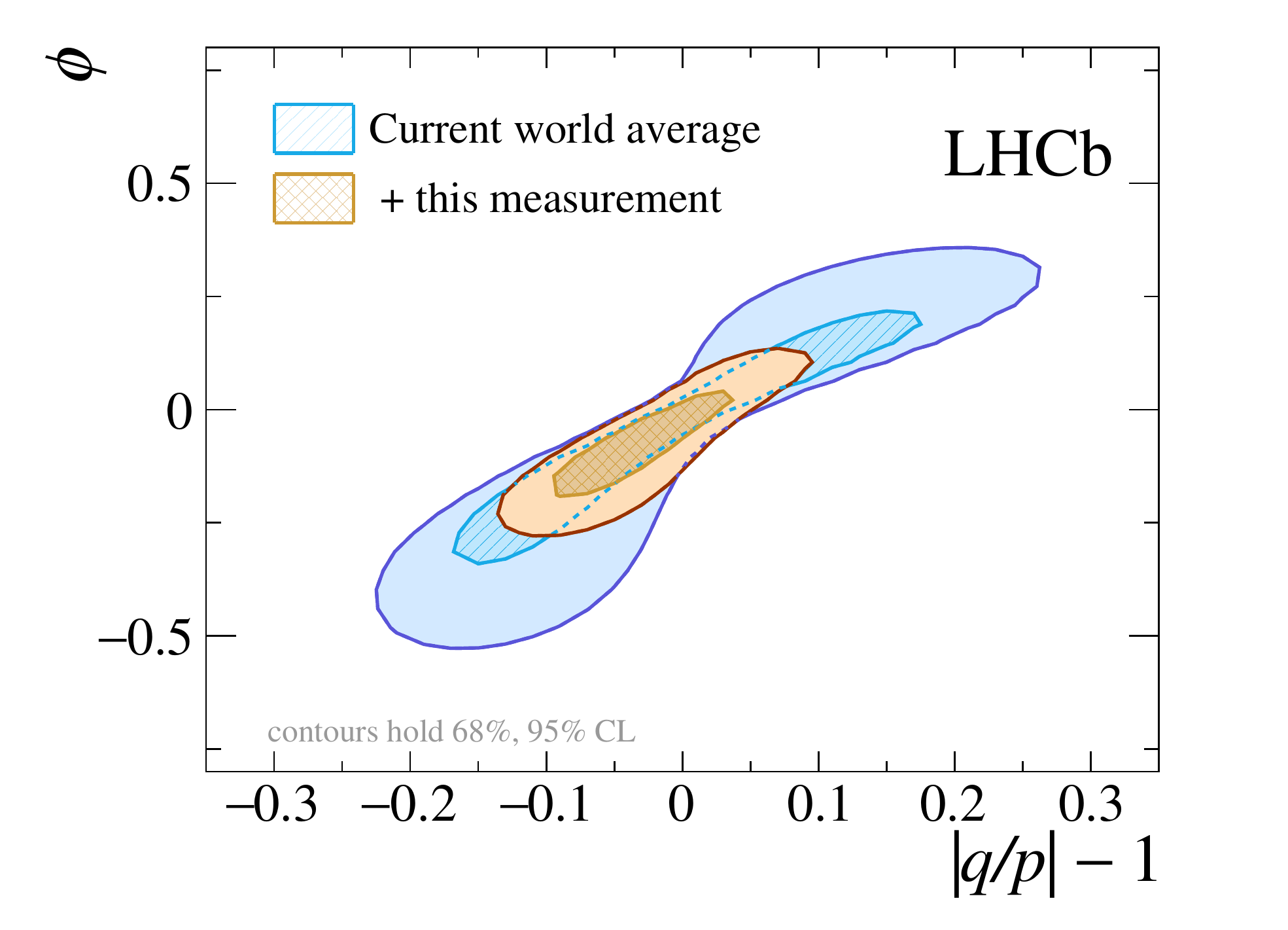}
    \vspace*{-0.5cm}
    \caption{
             Impact of the measurement~\cite{LHCb-PAPER-2019-001}, presented in Sect.~\ref{sect:D02KsPP}, on the previous world averages of (top) charm mixing parameters and (bottom) charm \CPV parameters.
            }
    \label{fig:D02KsPP_results}
\end{figure}

This is the most precise measurement of the parameter $x$ from a single experiment.
The impact of the results on the world average of the charm mixing and \CPV parameters is displayed in Fig.~\ref{fig:D02KsPP_results}.
In particular, the new world average gives the first evidence that $x>0$, \textit{i.e.} the mass of the \CP-even eigenstate of the charm neutral mesons is heavier than the \CP-odd one, at the level of $3\sigma$.

\section{Measurement of the $\vb*{CP}$ violation parameter $\vb*{A_\Gamma}$ with $\boldmath{\decay{\Dz}{\Kp\Km}}$ and $\boldmath{\decay{\Dz}{\pip\pim}}$ decays \label{sect:agamma}}

The time-dependent asymmetry between the decay rates of \Dz and \Dzb mesons into Cabibbo-suppressed final state $f = \Kp\Km$ or $f = \pip\pim$ can be written as
\begin{equation}
\label{eq:asymm}
\begin{aligned}
A_\CP(f,t) &\equiv \frac{\Gamma(\decay{\Dz}{f,t}) - \Gamma(\decay{\Dzb}{f,t})}
                   {\Gamma(\decay{\Dz}{f,t}) + \Gamma(\decay{\Dzb}{f,t})} \\
           &\approx A_\CP^\text{decay}(f) - A_\Gamma(f) \frac{t}{\tau_\Dz},
\end{aligned}
\end{equation}
where $A_\CP^\text{decay}$ is the \CP asymmetry in the decay, terms of order higher than one in time are neglected since the mixing parameters $x,y$ are both $<10^{-2}$~\cite{HFLAV} and the parameter $\agamma(f)$
is equal to~\cite{LHCb-CONF-2019-001}
\begin{equation}
\label{eq:agamma}
A_\Gamma(f) \approx - x\phi_f + y(\lvert q/p \rvert - 1) - y A_\CP^\text{decay}(f).
\end{equation}
Here, $\phi_f$ is defined as $\phi_f \equiv \arg[-(q A_f ) / ( p \bar{A}_f)]$, where $A_f$ ($\bar{A}_f$) is the decay amplitude of a \Dz (\Dzb) into final state $f$, and terms of order higher than one in the \CPV parameters $\phi_f$, $(\lvert q/p \rvert -1)$ and $A_\CP^\text{decay}$ are neglected~\cite{LHCb-CONF-2019-001}.
The last term in Eq.~(\ref{eq:agamma}) gives a contribution of order of $1\times 10^{5}$~\cite{LHCb-PAPER-2019-006,LHCb-PAPER-2016-035} and can be neglected at the current level of precision, approximately $3\times 10^{-4}$.
A nonzero measured value of \agamma would thus be a clear indication of \CPV in the mixing ($\lvert q/p \rvert \neq 1$) or in the interference between mixing and decay ($\phi_f \neq 0$).
Neglecting non-tree-level diagrams in the decay, which give contributions of order $\lvert (\Vub \Vcb) / (\Vus \Vcs) \rvert \approx 10^{-3}$~\cite{Grossman,Du}, the angle $\phi_f$ is equal to $\phi \equiv \arg(q/p)$ and \agamma is independent of the final state.
Under this approximation, both \CPV in the mixing and in the interference can be described by a single parameter, leading to the relation $x(\lvert q/p \rvert - 1) = y \phi$~\cite{indirect-CPV1,indirect-CPV2,indirect-CPV3}, which can be used to overconstrain the global fits to the charm \CPV parameters~\cite{HFLAV}.

Since the most recent SM predictions for \agamma are about $3\times 10^{-5}$~\cite{LHC-yellow},
measuring a nonzero value of \agamma at the current level of precision would indicate the presence of new phenomena beyond the SM.
In addition, increasing the precision of the measurement of \agamma is also important to measure the value of the \CPV in the decay from the measurements of the time-integrated asymmetries, as can be seen from Eq.~(\ref{eq:asymm})~\cite{LHCb-CONF-2019-001}.

The new measurement of \agamma presented at this conference by \lhcb uses the data sample collected during 2015--2016, corresponding to an integrated luminosity of $1.9\invfb$ of \proton\proton collisions at $\sqs = 13\tev$~\cite{LHCb-CONF-2019-001}.
The flavour at production of the \Dz is inferred from the charge of the pion in the strong decay $\decay{\Dstarp}{\Dz\pip}$, where the \Dstarp meson is produced in the \proton\proton primary vertex.
The candidates yield is 17 (5) million decays for the $\Kp\Km$ ($\pip\pim$) final state.
The raw time-dependent asymmetry between the measured yields of \Dz and \Dzb decays is equal to
\[
A_\text{raw}(f,t) \approx A_\CP(f,t) + A_\text{D}(\pip) + A_\text{P}(\Dstarp),
\]
where $A_\text{D}$ and $A_\text{P}$ indicate the detection and the production asymmetry, respectively, and terms of order three or higher in the asymmetries are neglected.
Owing to correlations between the momentum and the decay time which are caused by the selection requirements, momentum-dependent detection asymmetries of the tagging \pip and possible momentum-dependent production asymmetries of the \Dstarp meson cause detector-induced time-dependent asymmetries that bias the measurement of \agamma by a quantity larger than its statistical uncertainty.
These asymmetries are corrected for weighting the 3D momentum distributions of \Dz and \Dzb candidates to their average.
As a side effect, this causes a dilution of the measured value of \agamma to 88\% of its true value.
The results are corrected to account for this scale factor, which is estimated through a data-driven approach.

\begin{figure}[bt]
    \centering
    \includegraphics[width=70mm]{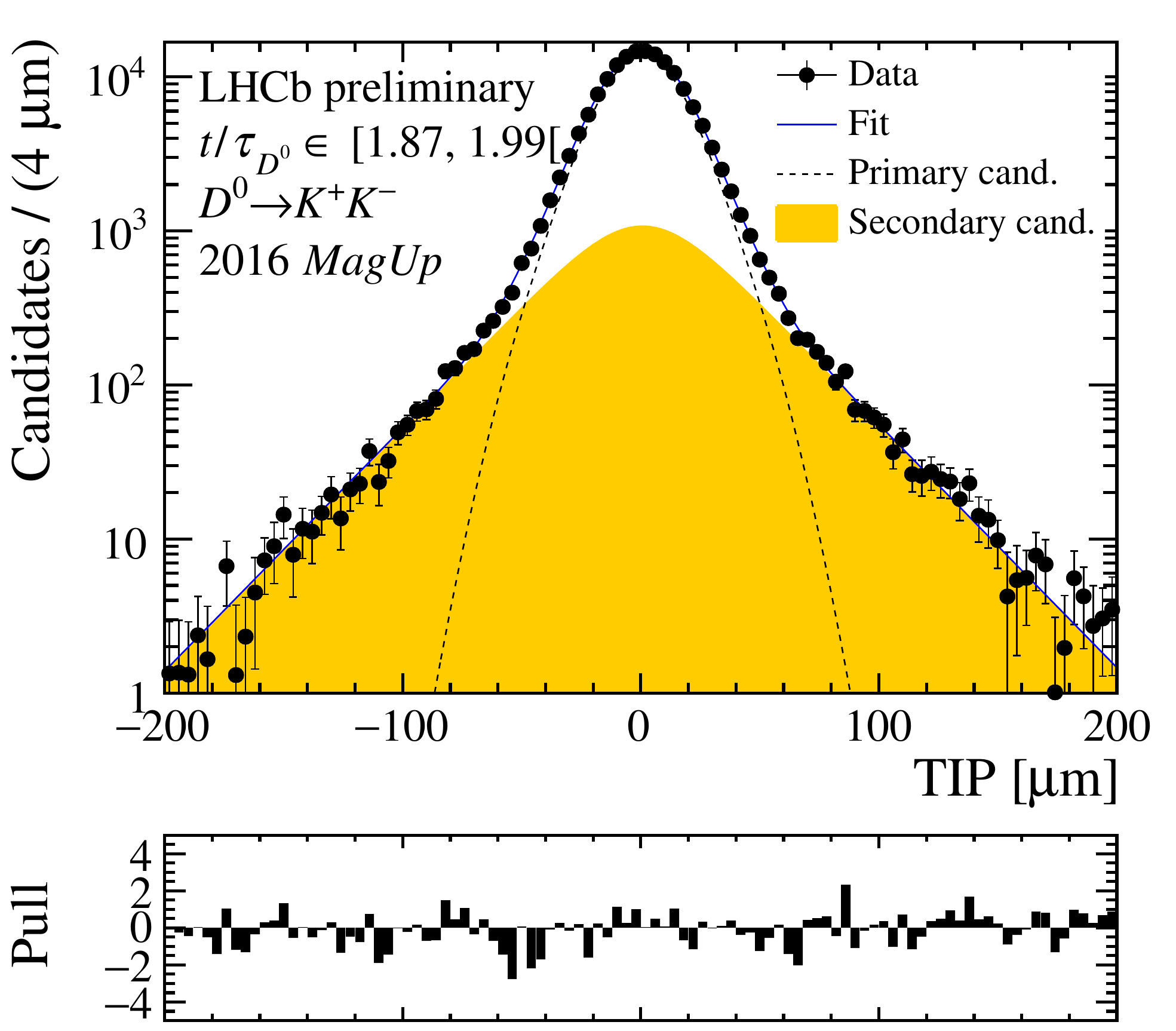}
    \vspace*{-0.2cm}
    \caption{
        Transverse IP distributions of \decay{\Dz}{\Kp\Km} decays for the \Dz candidates in the $11^{\text{th}}$ bin of decay time, for the events collected in 2016 with the dipole magnet polarity pointing upwards only.
        The results of the fits are superimposed, and the normalised residuals are displayed on the bottom part of the plot.
    }
    \label{fig:tip_fit}
\end{figure}
Secondary decays, where the \Dstarp meson is produced in the decay of a \bquark hadron, are a dangerous background since their production asymmetry is different from that of primary decays and their fraction in the data sample increases as a function of time up to values greater than 10\%.
As a consequence, they would bias the measurement of \agamma if they were unaccounted for in the measurement of the asymmetry.
The asymmetry of primary decays is disentangled from that of secondary decays through fits to the distributions of the impact parameter in the plane transverse to the beam (TIP), simultaneously in 21 bins of decay time in the range $[0.6,8]\tau_\Dz$ and for \Dz and \Dzb candidates.
Following simulation studies, in each decay time bin an exponential function is taken as the functional shape for the TIP distribution of secondary decays and the resolution function, which is shared by primary and secondary decays, is parametrised as the sum of two Gaussian functions whose width increases as an erf function as a function of decay time.
An example of the fits is displayed in Fig.~\ref{fig:tip_fit}.
Finally, a linear function is fitted to the so-obtained time-dependent asymmetry of primary decays to measure \agamma.
The analysis procedure is validated using the Cabibbo-favoured \decay{\Dz}{\Km\pip} decays, for which the analogue $\agamma^{\kaon\pi}$ of the parameter \agamma is negligible compared to the experimental uncertainty.
The result is $\agamma^{\kaon\pi} = (0.7 \pm 1.1)\times 10^{-4}$, where only the statistical uncertainty is considered.

The main systematic uncertainty on the measurement is due to the uncertainty on the knowledge of the fraction of secondary decays at low decay times.
This is evaluated repeating the TIP fits with different resolution functions and constraining the fraction of secondary decays to that obtained in simulation or inferred using a sample that combines \Dstarp with \mun candidates, which is highly enriched in secondary decays.
Thanks to the improved analysis procedure, this uncertainty, which corresponds to $0.4\times 10^{-4}$, is more than halved with respect to that of the previous \lhcb measurement with 2011--2012 data~\cite{LHCb-PAPER-2016-063}.
Further contributions to the systematic uncertainties are due to residual backgrounds under the \Dz mass peak from partially-reconstructed \Dz multibody decays, to possible correlations of the background asymmetry with the value of the reconstructed mass of the \Dstarp meson, and to the choice of the binning for the kinematic weighting.
These uncertainties are all of order of $0.3 \times 10^{-4}$, and the estimate of the last two is dominated by the size of the available data sample.

\begin{figure}[bt]
    \centering
    \includegraphics[width=80mm]{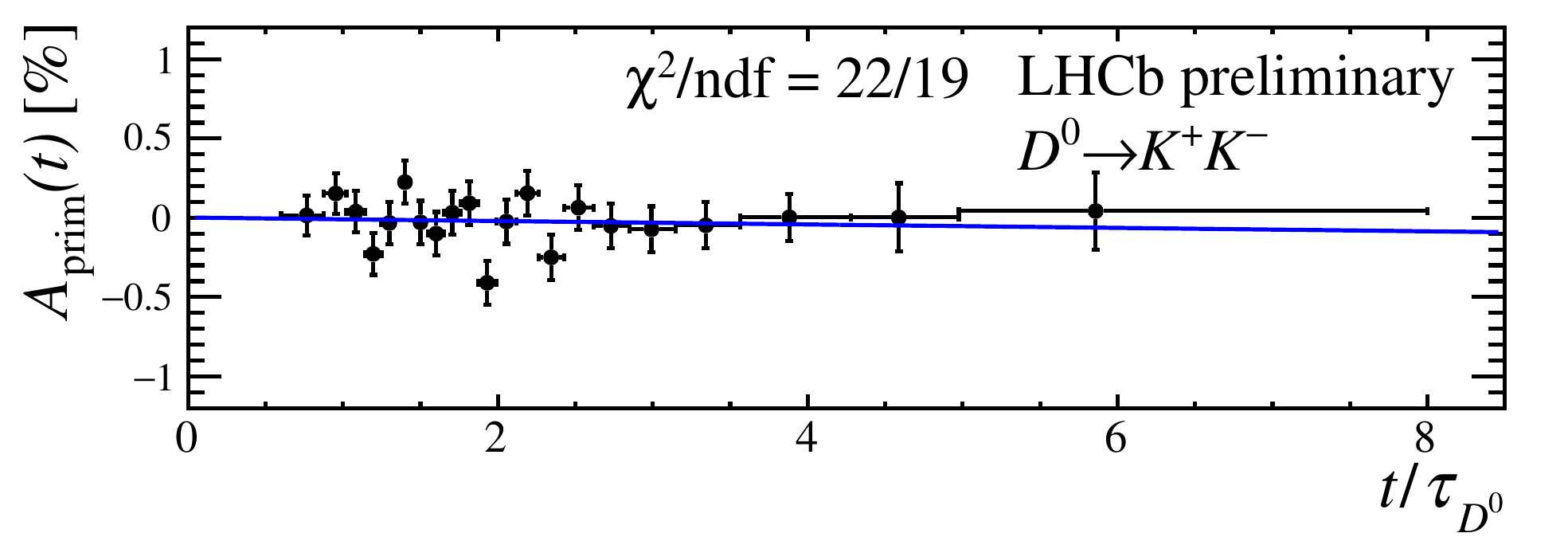}
    \vspace*{-0.4cm}
    \includegraphics[width=80mm]{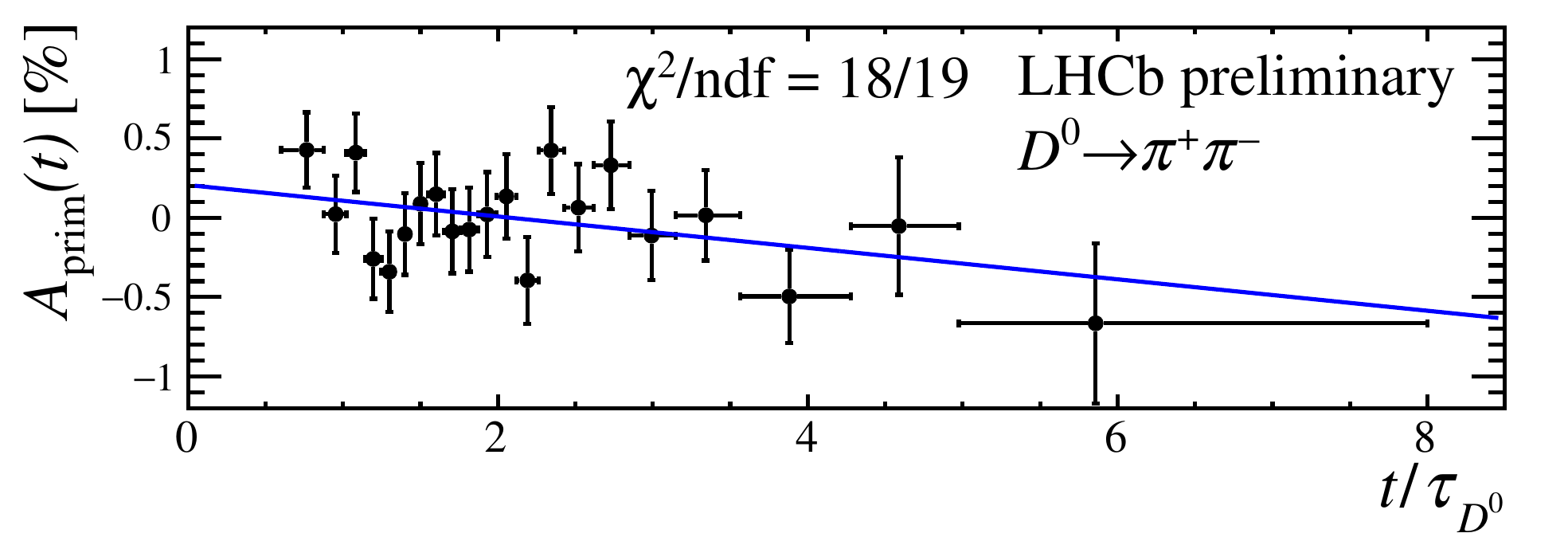}
    \caption{
        Fitted asymmetry of the primary decays in bins of decay time expressed in units of \Dz lifetimes, $\tau_\Dz = 0.410\ps$~\cite{PDG2018}, for (top) the \decay{\Dz}{\Kp\Km} and (bottom) the \decay{\Dz}{\pip\pim} decay channels.
        The solid lines show the linear fit, whose slope is equal to $-\agamma$.
    }
    \label{fig:agamma_fit}
\end{figure}
The fits to the time-dependent asymmetry are displayed for both decay channels, after the kinematic weighting and the subtraction of secondary decays, in Fig.~\ref{fig:agamma_fit}.
The measured values of \agamma are
\begin{align*}
A_\Gamma(\Kp\Km)   &= (\phantom{1}1.3 \pm 3.5 \pm 0.7) \times 10^{-4}, \\
A_\Gamma(\pip\pim) &= (          11.3 \pm 6.9 \pm 0.8) \times 10^{-4},
\end{align*}
where the first uncertainties are statistical and the second systematic.
Neglecting weak phases in the decay~\cite{Grossman,Du}, the two values can be combined with each other and with the results of the previous measurement from \lhcb that used \Dstarp-tagged data collected during 2011--2012~\cite{LHCb-PAPER-2016-063}, obtaining
\[
\agamma(\Kp\Km + \pip\pim) = (0.9 \pm 2.1 \pm 0.7) \times 10^{-4}.
\]
This result is compatible with the hypothesis of no \CPV and dominates the uncertainty of the world average of \agamma~\cite{HFLAV}.

\section{Conclusions and prospects \label{sect:conclusions}}
The \lhcb experiment has provided the most precise measurements to date of the parameters of mixing and time-dependent \CPV in charm thanks to the huge data sample of charm decays collected and to a tight control of the systematic uncertainties.
In these two specific fields, the achieved precision is already comparable or even better that that foreseen by the Belle~II collaboration at the end of the Belle~II data taking~\cite{Belle2}.

While there is evidence that both the width and the mass split of the charm neutral mesons are nonzero, the SM predictions for time-dependent \CPV are still one order of magnitude below the current experimental precision.
In the long term, the Upgrade~I (2021--2029)~\cite{UpgradeI} and the proposed Upgrade~II (2031--2038)~\cite{UpgradeII} of \lhcb will be essential to test these predictions.
However, a significant increase in precision is expected already in the very near future thanks to the steady increase of the size of the \lhcb collected data sample over the past few years and to the new measurements of the strong phases of \Dz-meson decays foreseen by the BESIII collaboration, which will help to keep the systematic uncertainty on the mixing parameters below the statistical one.
For example, about 30 times more \decay{\Dz}{\KS\pip\pim} decays were collected by \lhcb during 2015--2018 with respect to those
of 2011--2012 that are exploited in the analysis presented in Sect.~\ref{sect:D02KsPP}.
As far as two-body decays are concerned, the data collected during 2017--2018 correspond to an integrated luminosity of about $3.8\invfb$, to be compared with $1.9\invfb$ collected during 2015--2016 and used in the measurement of \agamma presented in Sect.~\ref{sect:agamma} and in the latest measurement of mixing and \CPV in \decay{\Dz}{\Kpm\pimp} decays~\cite{LHCb-PAPER-2017-046}.


\bigskip 

\end{document}